\documentclass[%
 prb,twocolumn,10pt,
superscriptaddress,
 amsmath,amssymb,
 aps,
prm,
]{revtex4-2}

\usepackage[english]{babel}

\usepackage[letterpaper,top=2cm,bottom=2cm,left=3cm,right=3cm,marginparwidth=1.75cm]{geometry}

\usepackage{amsmath}
\usepackage{graphicx}
\usepackage[colorlinks=true, allcolors=blue]{hyperref}
\usepackage{xcolor}
\usepackage{ulem}

\begin{document}

\preprint{APS/123-QED}

\title{Synthetic control over marcasite-pyrite polymorph formation in the Fe$_{1-x}$Co$_x$Se$_2$ series}

\author{Luqman Mustafa}
\affiliation{Experimental Physics IV, Ruhr University Bochum, Universitätsstraße 150, 44801 Bochum, Germany}

\author{Susanne Kunzmann}
\affiliation{Applied Quantum Materials, Institute for Energy and Materials Processes (EMPI),  Faculty of Engineering, University of Duisburg-Essen, Forsthausweg 2, 47057 Duisburg, Germany}
\affiliation{Interdisciplinary Centre For Advanced Materials Simulation (ICAMS), Ruhr University Bochum, Universitätsstraße 150, 44801 Bochum, Germany}
\affiliation{Research Center Future Energy Materials and Systems (RC FEMS), University of Duisburg-Essen,  Forsthausweg 2, 47057 Duisburg, Germany}

\author{Martin Kostka}
\affiliation{Experimental Physics IV, Ruhr University Bochum, Universitätsstraße 150, 44801 Bochum, Germany}

\author{Jill Fortmann}
\affiliation{Institute for Materials, Ruhr University Bochum, Universitätsstraße 150, 44801 Bochum, Germany}

\author{Aurelija Mockute}
\affiliation{Institute for Materials, Ruhr University Bochum, Universitätsstraße 150, 44801 Bochum, Germany}

\author{Alan Savan}
\affiliation{Institute for Materials, Ruhr University Bochum, Universitätsstraße 150, 44801 Bochum, Germany}

\author{Alfred Ludwig}
\affiliation{Institute for Materials, Ruhr University Bochum, Universitätsstraße 150, 44801 Bochum, Germany}
\affiliation{Center for Interface-Dominated High Performance Materials (ZGH), Ruhr University Bochum, Universitätsstraße 150, 44801 Bochum, Germany}
\affiliation{Research Center Future Energy Materials and Systems (RC FEMS), Ruhr University Bochum, Universitätsstraße 150, 44801 Bochum, Germany}

\author{Anna Grünebohm}
\affiliation{Interdisciplinary Centre For Advanced Materials Simulation (ICAMS), Ruhr University Bochum, Universitätsstraße 150, 44801 Bochum, Germany}
\affiliation{Faculty for Physics and Astronomy, Ruhr University Bochum, Universitätsstr. 150, 44801 Bochum, Germany}
\affiliation{Center for Interface-Dominated High Performance Materials (ZGH), Ruhr University Bochum, Universitätsstraße 150, 44801 Bochum, Germany}

\author{Andreas Kreyssig}
\affiliation{Experimental Physics IV, Ruhr University Bochum, Universitätsstraße 150, 44801 Bochum, Germany}

\author{Anna E. Böhmer}
\affiliation{Experimental Physics IV, Ruhr University Bochum, Universitätsstraße 150, 44801 Bochum, Germany}

\begin{abstract}
Transition-metal dichalcogenides of the pyrite-marcasite family are model systems of crystal chemistry. A few of these show polymorphism. The theoretical ground state of CoSe$_2$ is marcasite, but the material is typically synthesized in the pyrite structure. Polymorphism has been observed in nanoparticles and synthetic control of the polymorphs of CoSe$_2$ has not been achieved. 
We have synthesized material libraries of the Fe$_{1-x}$Co$_x$Se$_2$ series by combining combinatorial deposition and ex-situ selenization. The approach allows to efficiently explore substitution ranges and crystal structures that form for different synthesis conditions. We find that higher levels of Co content $x$ within the marcasite structure are possible when synthesizing at low temperatures. At a synthesis temperature of only 250\textdegree\,C, we have successfully synthesized marcasite CoSe$_2$ as the majority phase.
Density functional theory simulations reveal that the two isomorphs of CoSe$_2$ are extremely close in energy and that the orthorhombic phase is the energetic ground state. 
Our experimental and theoretical data show that the marcasite structure is the equilibrium phase of Fe$_{1-x}$Co$_x$Se$_2$ in the entire composition range.  
\end{abstract}

\maketitle

\section{Introduction}

Transition-metal dichalcogenide compounds of the pyrite-marcasite family have attracted significant attention as long-standing model systems for crystal chemistry\cite{Buerger1937,Kjekshus1974-1, Kjekshus1974-2, Kjekshus1975, Witthaut2025} and for their potential applications. Specifically, FeSe$_2$ and CoSe$_2$ show favorable catalytic properties for the hydrogen evolution reaction and the oxygen reduction reaction \cite{Zhang2015,Kim2023,Sureshkumar2025}. Additionally, FeSe$_2$ is a semiconductor with a narrow bandgap, making it useful for applications such as photodetectors, solar cells, and light-emitting diodes. Co substitution in FeSe$_2$ increases its conductivity, which has also been pointed out as an approach to increase the thermoelectric performance \cite{Kim2023}. 

 Only a few compounds are known to show polymorphism and exist in both the pyrite and the marcasite structure \cite{Kjekshus1975}. Of these polymorphic compounds, the most common compound is pyrite, FeS$_2$, from which the cubic structure derives its mineral name. However, FeS$_2$ can also naturally occur in the orthorhombic marcasite structure, a derivative of the cubic pyrite structure. 
 Synthetic control over the polymorphs of FeS$_2$ has only recently been achieved \cite{Ma2021} by space-separated hydrothermal synthesis. 
 
Bulk CoSe$_2$ commonly exists in the cubic pyrite structure. Nevertheless, density functional theory (DFT) calculations predict an orthorhombic marcasite ground state  \cite{Gavhane_2021, Zhang2019}. A mineral sample of orthorhombic marcasite CoSe$_2$ is known \cite{Ramdohr1955}, which constitutes the structural reference  still to date. In nanoparticles of CoSe$_2$, polymorphism of both orthorhombic marcasite and cubic pyrite structures has been observed and was connected to enhanced catalytic properties \cite{Zhang2015, Ye_2022, Zhang_2023}. It seems that the orthorhombic material has a better performance than the cubic material \cite{Ye_2022}. For the substitution series Fe$_{1-x}$Co$_x$Se$_2$, it has been reported that the cubic structure prevails when the Co content is 75\% or higher, and the orthorhombic structure prevails for lower Co contents \cite{Kim2023}. Bulk FeSe$_2$ indeed always forms in the orthorhombic structure.

Here, we investigate the synthesis and polymorphism of the whole Fe$_{1-x}$Co$_x$Se$_2$ series. We use a two-step synthesis approach, consisting of the deposition of a thin-film Fe-Co materials library\cite{Ludwig2019} and ex-situ selenization, which permits a high level of control of the chemical composition and synthesis temperature. 
We show that the synthesis temperature is crucial for the stabilization of the polymorphs. DFT analysis shows that the two polymorphs of CoSe$_2$ are extremely close in energy, highlighting that the compound is at the brink of a crystallographic transition. The energy difference between the two polymorphic phases is systematically increased by Fe-substitution.   
Indeed, we find that the orthorhombic structure is stabilized in Fe$_{1-x}$Co$_x$Se$_2$ to higher Co content when the synthesis temperature is low. At the lowest synthesis temperature of 250\textdegree\,C, the majority phase is orthorhombic CoSe$_2$, showing synthetic control of the polymorphic phases of CoSe$_2$. 

\section{Methods}
Fe$_{1-x}$Co$_x$Se$_2$ films were synthesized via a two-step process, similar to Ref.~\cite{Kostka2025}. In the first step, a thin-film materials library\cite{Ludwig2019} having a continuous Fe-Co gradient with a thickness of approximately 200\,nm was deposited at room temperature onto thermally oxidized silicon strips (70--100\,mm long, 6\,mm wide, 525\,\textmu m thick, with a 2\,\textmu m SiO$_2$ buffer layer), using magnetron co-sputtering. High-purity Fe (99.95\%) and Co (99.97\%) targets (1.5-inch diameter, 0.125-inch thick, Lesker) were sputtered in an high purity argon atmosphere (99.9997\%) at a pressure of $5 \times10^{-2}$\,mbar with radio frequency power source of 30~W for Fe and 32~W for Co. The Fe and Co targets were positioned 180\textdegree{} apart at an angle of 77.5\textdegree{} relative to the substrate plane, resulting in an almost linear composition gradient from Fe-rich to Co-rich across the thin-film materials library [see Fig. 1(a)]. Pure Fe and Co films were also prepared under similar conditions.

In the second, ex-situ step [see Fig. 1 (b,c)], the material libraries were sealed in evacuated quartz tubes (vacuum better than $8 \times 10^{-3}$\,mbar) together with 50\,mg of high-purity Se (99.999\%). The tubes were placed in a box furnace with the material library facing down to minimize Se droplet deposition, and heated to synthesis temperatures, $T_\mathrm{synthesis}$, of 250\textdegree\,C, 350\textdegree\,C and 430\textdegree\,C at a rate of 100\textdegree\,C/h. The temperature was held for 12 -- 60\,hours, after which the samples were cooled to room temperature at a rate of 100\textdegree\,C/h.

\begin{figure}
    \centering
    \includegraphics[width=\columnwidth]{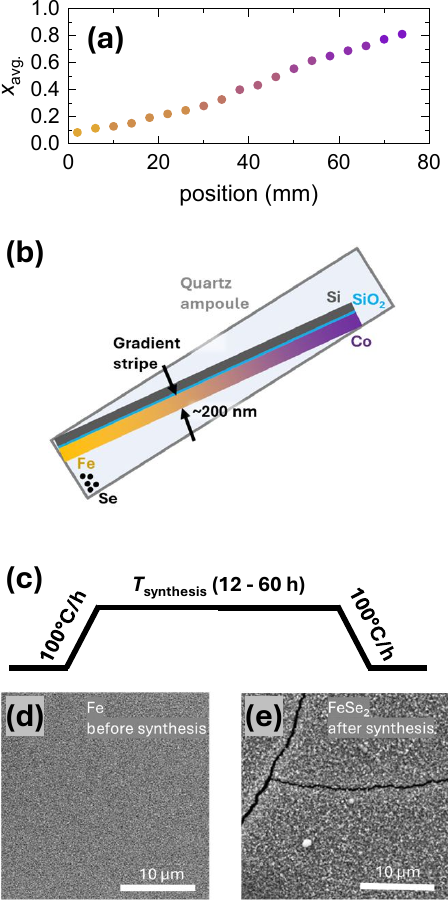}
    \caption{Synthesis of Fe$_{1-x}$Co$_x$Se$_2$ thin-film materials libraries (a) Variation of the relative Co content $x_\mathrm{avg.}$ along the materials library as determined by EDS. (b) Sketch of the materials library enclosed in a quartz tube for selenization, not to scale. (c) Temperature profile of the selenization process. (d,e) SEM images of a materials library before and after the selenization process, respectively.}
    \label{Fig:1}
\end{figure}

The films' morphology was evaluated in a FEI Quanta SEM, before and after the selenization process. The transition-metal films before selenization are smooth with a closed surface. The selenized films are granular and remain mostly closed, with some cracks forming, most likely due to the expansion of film's volume during selenization [see Fig. 1 (d,e)]. The films' composition were characterized using energy-dispersive X-ray spectroscopy (EDS) at different positions on the substrate, each 4 mm apart. The relative Co content $x_\textnormal{avg.}$ is quantified as

\begin{equation}
    x_\textnormal{avg.}=\frac{\mathrm{at}\%\mathrm{Co}}{\mathrm{at}\%\mathrm{Fe}+\mathrm{at}\%\mathrm{Co}}
\end{equation}

As the films may contain multiple crystallographic phases, $x_\textnormal{avg.}$ is an average value of 
Co-content in all phases present. Finally, the structural properties of these films were investigated by X-ray diffraction (XRD) using a Bruker D8 with high-brilliance IµS microfocus X-ray source (Cu) and VÅNTEC 500 – 2D detector. The measurements were performed at the same positions on the films as the EDS measurements and the resulting patterns were refined using TOPAS. 

DFT simulations were performed with the abinit package \cite{gonze_recent_2016} using the PBE \cite{Perdew_96} optimized norm conserving pseudopoentials from PSEUDODOJO \cite{VANSETTEN201839} with $3$s$^23$p$^64$s$^2
3$d$^7$, $3$s$^23$p$^64$s$^23$d$^6$ and $4$s$^23$d$^{10}4$p$^4$ electrons treated as valence for Co, Fe and Se, respectively. 
Gaussian smearing (0.27~eV) together with a $8\times 8 \times 8$ $k$-mesh  for the cubic phase (rescaled to $8\times 8 \times 12$ for the orthorhombic lattice) and a plane-wave cutoff of 1496.63~eV  guarantee energy convergence of about 1~meV. Volumes and ionic positions were optimized until the maximal forces were below $2.5\cdot10^{-3}$ eV/\AA{} and the  threshold for self-consistency was set to  $2.72\cdot 10^{-7}$~eV. In agreement to literature \cite{Gavhane_2021} we found no stable magnetic moments for CoSe$_2$ and only results from non-magnetic simulations are presented.
For the substitution of Fe by Co, all possible lattice sites in the unit cell (cubic) or in $2\times 1 \times 1$ supercells (orthorhombic) were considered and averaged values for the energies and lattice constants were used.

\section{Experimental results and analysis}

\begin{figure*}
    \centering
    \includegraphics[width=\textwidth]{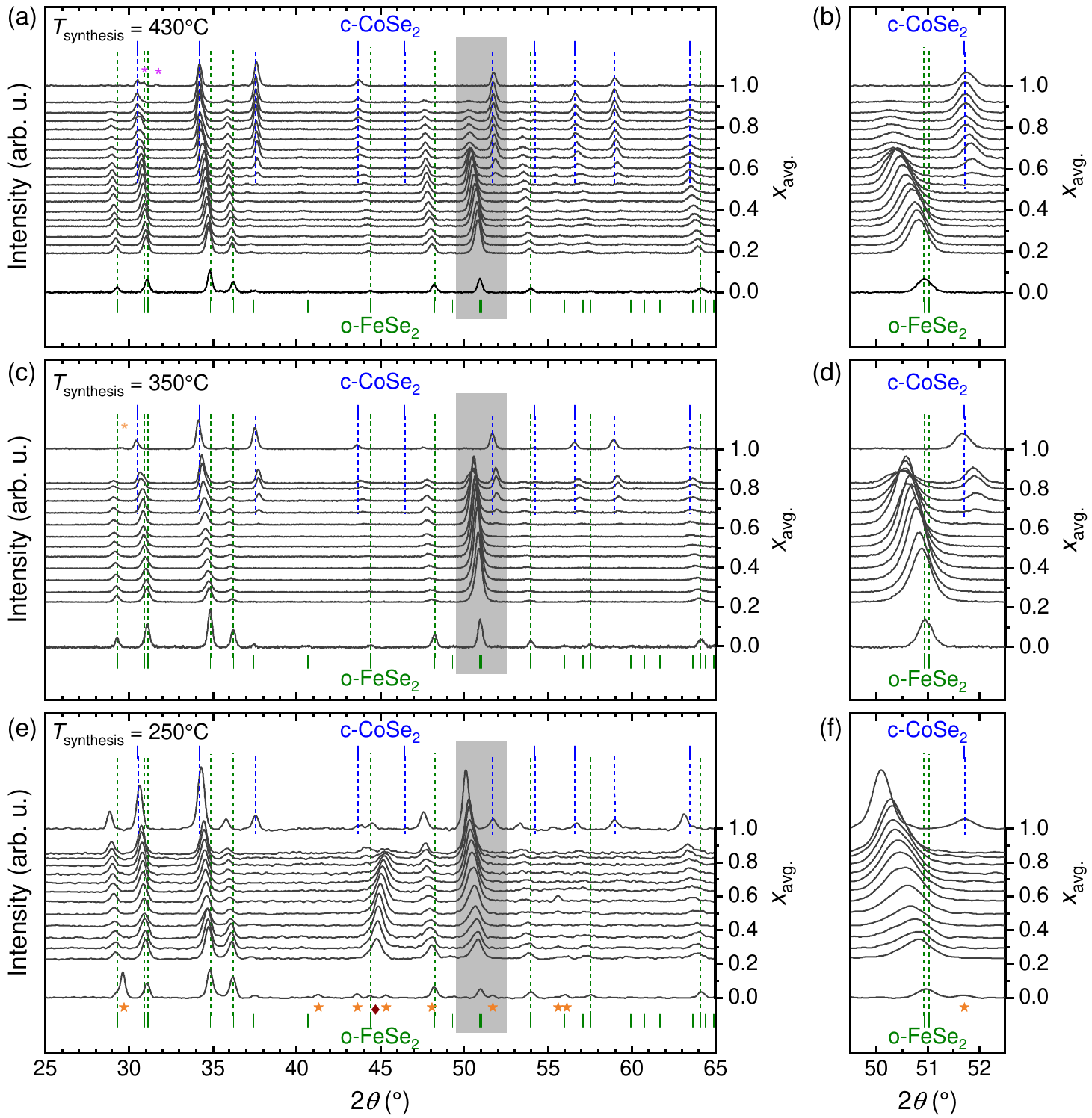}
    \caption{XRD patterns for Fe$_{1-x}$Co$_x$Se$_2$ thin-film materials library measured in 4 mm steps. Data for films of pure Fe and Co are added. The data are stacked according to $x_\textnormal{avg.}$ for clarity, with the scale indicated on the right axis. (a),(b) XRD patterns for the films synthesized at 430\textdegree C. (c),(d) XRD patterns for the film synthesized at 350\textdegree C. (e),(f) XRD patterns for the films synthesized at 250\textdegree C. Green ticks and lines represent the Bragg peak positions of the orthorhombic marcasite o-FeSe$_2$ phase. The blue ticks and lines represent the Bragg peak positions for the cubic pyrite c-CoSe$_2$ phase. Orange stars show the Bragg peak positions of elemental Se. In panel (e), the closed diamond indicates the (101) peak position of elemental Fe and the open diamond indicates the position of (111) Bragg peak of elemental Co.}
    \label{Fig:2}
\end{figure*}

\begin{figure*}
    \centering
    \includegraphics[width=\textwidth]{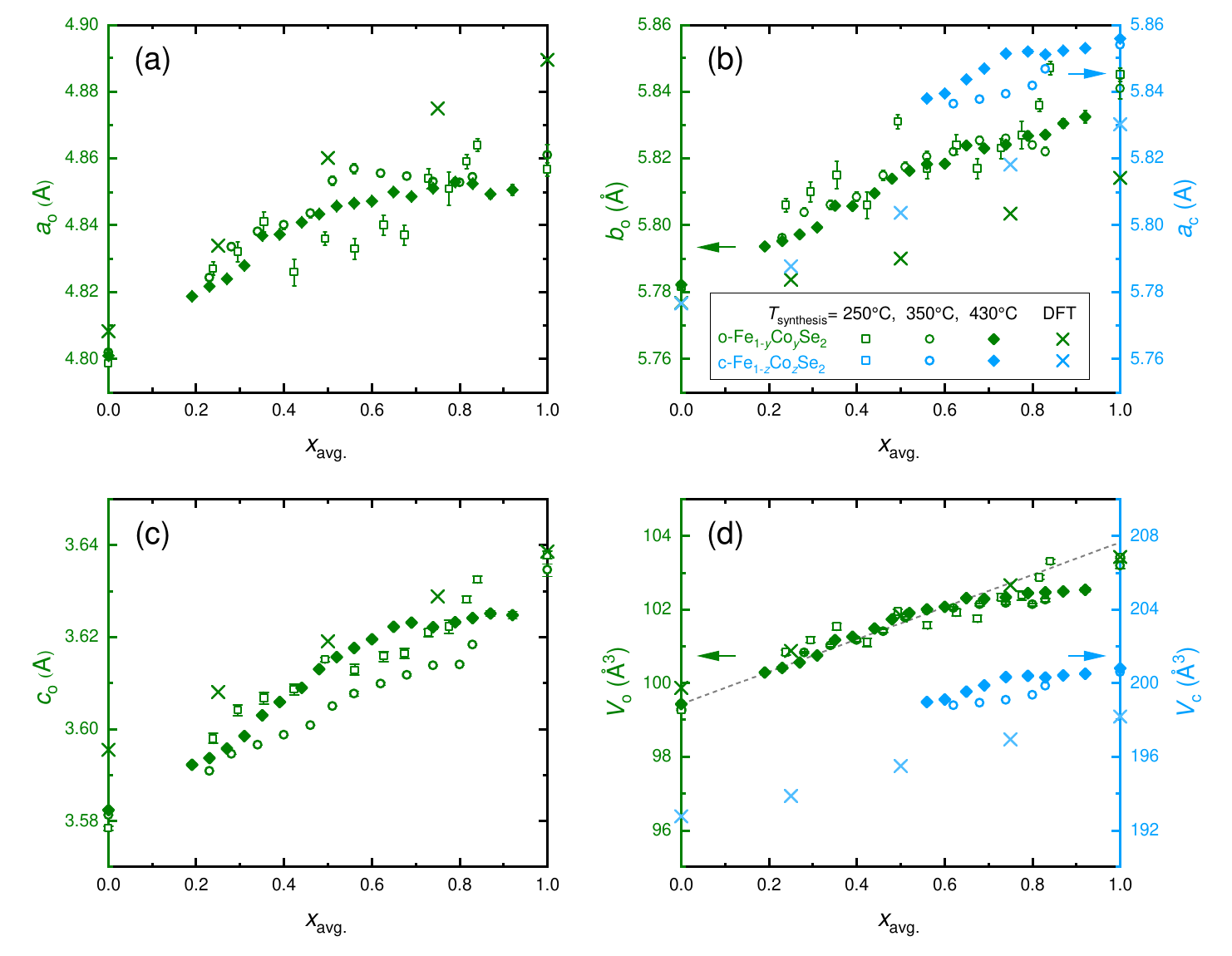}
    \caption{Structural parameters of o-Fe$_{1-x}$Co$_x$Se$_2$ and c-Fe$_{1-x}$Co$_x$Se$_2$ as a function of $x_\textnormal{avg.}$ for different synthesis temperatures. Experimental and DFT orthorhombic $a_{\textnormal{o}}$ lattice parameter (a), orthorhombic $b_{\textnormal{o}}$ lattice parameter and cubic $a_{\textnormal{c}}$ lattice parameter (b), orthorhombic $c_{\textnormal{o}}$ lattice parameter (c), as well as orthorhombic and cubic unit-cell volumes, $V_{\textnormal{o}}$ and $V_{\textnormal{c}}$ (d). Note that the cubic unit cell contains twice as many formula units as the orthorhombic unit cell. 
    }
    \label{Fig:3}
\end{figure*}

The Fe$_{1-x}$Co$_x$Se$_2$ films were characterized in 4 mm steps by SEM and XRD measurements. The EDS measurements confirm a linear composition gradient on the film with $x_\textnormal{avg.}$ typically ranging from 10--80\% as in Fig.~\ref{Fig:1}~(a). The ratio of (Fe+Co) to Se is in most cases 1:2 in the uncertainty range of the EDS measurement.
 
Figure~\ref{Fig:2} shows the XRD patterns of the films selenized at different $T_\mathrm{synthesis}$. The peak positions for pure orthorhombic FeSe$_2$ and pure cubic CoSe$_2$ are indicated. Typically, a gradient film covers compositions of $x_\textnormal{avg.}\sim0.2-0.9$. Unary films of elemental Fe and Co were selenized additionally. We describe all XRD patterns by a combination of the orthorhombic marcasite o-Fe$_{1-y}$Co$_y$Se$_2$ and cubic pyrite c-Fe$_z$Co$_{1-z}$Se$_2$ phases with modified lattice parameters. Note that the Co content of the individual phases $y$, $z$ is a priori unknown when more than one phase is present at the same position, as only the average Co content $x_\textnormal{avg.}$ can be determined directly via EDS. 

The films synthesized at $T_\mathrm{synthesis}=430$\textdegree\,C crystallized in the orthorhombic o-Fe$_{1-y}$Co$_y$Se$_2$ structure for Co content lower than $x_\textnormal{avg.}<0.56$. The position of the Bragg peaks shift to lower $2\theta$ with increasing average Co content $x_\textnormal{avg.}$, showing an increase of the lattice parameters. At $x_\textnormal{avg.}=0.56$, additional Bragg peaks, which are attributed to the cubic c-Fe$_z$Co$_{1-z}$Se$_2$ phase, start to appear. Further increases in the average Co content result in an increase in the intensity of the associated peaks.  Their position shifts towards higher $2\theta$ which, similarly, indicates a change of the lattice parameters. The films synthesized at $\textit{T}_\mathrm{synthesis}=350$\textdegree C show qualitatively the same behavior as the films synthesized at $\textit{T}_\mathrm{synthesis}=430$\textdegree C. However, the cubic phase starts to appear at slightly higher Co content of $x_\textnormal{avg.}\geq 0.62$.

In the film synthesized at the lowest temperature of  $T_\mathrm{synthesis}=250$\textdegree\,C, the cubic phase is not observed except for the extreme composition $x_\textnormal{avg.}=1$ and, even at this value, the majority of the film still crystallized in the orthorhombic structure o-CoSe$_2$. Note that at such a low synthesis temperature, a significant fraction of the Fe/Co remained unreacted with Se when the synthesis was performed over 12 hours, resulting in additional XRD peaks of elemental Fe and Co. This unreacted fraction was much reduced by increasing the selenization time from 12 to 60 hours for the extreme compositions FeSe$_2$ and CoSe$_2$. 

\begin{figure}
    \centering
    \includegraphics[width=\columnwidth]{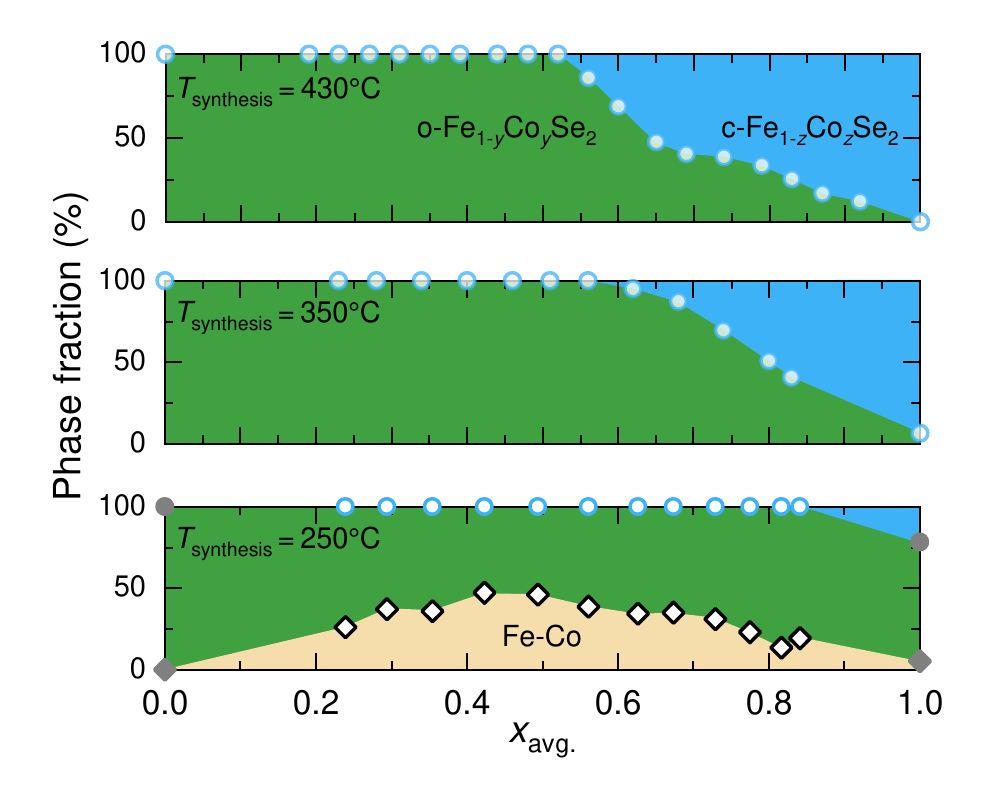}
    \caption{Phase fraction as function of the experimentally determined average Co content $x_\textnormal{avg.}$ for films synthesized at different temperatures. The green area represents the orthorhombic phase fraction and blue area represents the cubic phase fraction. The beige area represents the unreacted fraction of Fe and Co. All films were selenized for 12 hours (open symbols) except for the pure iron and cobalt at 250\textdegree C (closed grey symbols), which were selenized for 60 hours.}
    \label{Fig:4-1}
\end{figure}

Rietveld refinements were performed to obtain the lattice parameters and phase fractions from the XRD patterns. In the refinement procedure, we considered the orthorhombic marcasite phase of FeSe$_2$ \cite{Kjekshus1974-1}, cubic pyrite and orthorhombic marcasite phases of CoSe$_2$ \cite{Furuseth1969Magnetic, Ramdohr1955}, as well as elemental phases of Se, Fe, and Co \cite{Cherin1967TrigonalSe, Woodward2003DoublePerovskite, Taylor1950LatticeParameters}. 
From the variation of lattice parameters, we deduce that FeSe$_2$ is partially substituted with Co to o-Fe$_{1-y}$Co$_y$Se$_2$ and CoSe$_2$ is partially substituted with Fe to c-Co$_{1-z}$Fe$_z$Se$_2$. Note that the average Co content, $x_\textnormal{avg.}$, as measured via EDS represents an average of all phases present in a film and that the individual Co contents $y$ and $z$ cannot be determined directly.

Figure~\ref{Fig:3} shows the evolution of the lattice parameters of the o-Fe$_{1-y}$Co$_y$Se$_2$ and c-Co$_{1-z}$Fe$_z$Se$_2$ phases with $x_\textnormal{avg.}$ of the three films with different $T_\mathrm{synthesis}$. Overall, all lattice parameters increase with increasing $x_\textnormal{avg.}$. The lattice parameters of o-Fe$_{1-y}$Co$_y$Se$_2$ appear to saturate around $x_\textnormal{avg.}=0.56$ and $x_\textnormal{avg.}=0.62$ for $T_\mathrm{synthesis}=430$\textdegree C and 350\textdegree C, respectively. This indicates an increasing difference between the Co content $y$ of the o-Fe$_{1-y}$Co$_y$Se$_2$ phase from $x_\textnormal{avg.}$. Despite a larger scatter in the data, this saturation is absent for the lowest $T_\mathrm{synthesis}=250$\textdegree C. The lattice parameter of the c-Fe$_{1-\textit{z}}$Co$_{\textit{z}}$Se$_2$ phase, $a_c$, is approximately constant between $x_\textnormal{avg.} = 0.75-1$ and decreases with further decreasing Co content, which is seen most clearly for $T_\mathrm{synthesis} = 430$\textdegree C.

Figure~\ref{Fig:4-1} shows the evolution of phase fractions with Co content and synthesis temperature as obtained from the refinement. For pure CoSe$_2$ with $x_\textnormal{avg.}=1$, the cubic phase dominates for synthesis at $T_\mathrm{synthesis}=350$\textdegree C and 430\textdegree C. In contrast, the orthorhombic o-CoSe$_2$ dominates in films selenized at $T_\mathrm{synthesis}=250$\textdegree C with only $~25$\% of the cubic phase present. For pure FeSe$_2$ only the orthorhombic phase is present for all synthesis temperatures. 

\begin{figure}
    \centering
    \includegraphics[width=\columnwidth]{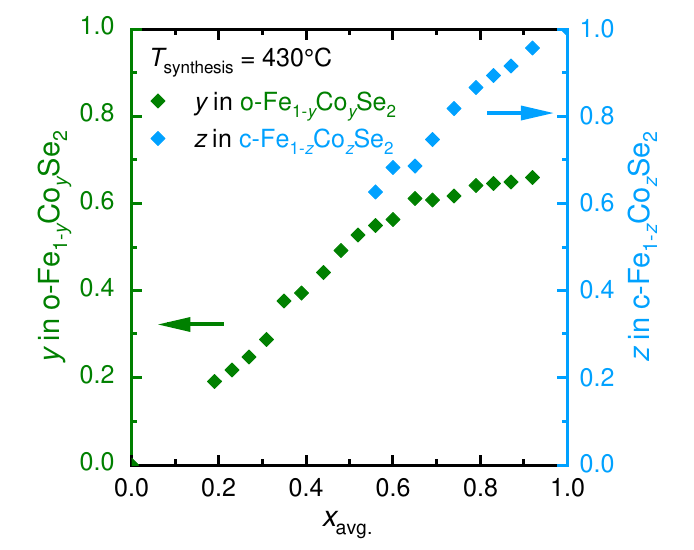}
    \caption{Estimated cobalt content $y$ in the orthorhombic marcasite phase, and cubic pyrite phase, $z$, as a function of $x_\mathrm{avg.}$ for the film selenized at 430\textdegree C, where the most accurate structural data were obtained.}
    \label{Fig:4-2}
\end{figure}

We can deduce the respective Co contents $y$ and $z$ of o-Fe$_{1-y}$Co$_y$Se$_2$ and c-Fe$_{1-z}$Co$_{z}$Se$_2$ in the phase coexistence range from our experimental data of the films selenized at 430\textdegree C. First, we assume a linear dependence of the unit cell volume of o-Fe$_{1-y}$Co$_y$Se$_2$ on Co content $y$, following Vegard's law [dashed line in Fig. \ref{Fig:3}(d)] as found for the pure phases in DFT simulations. Second, we assume that only the two phases are present so that $x_\textnormal{avg.}$ is an average of $y$ and $z$ weighted by the respective phase fractions. Figure\ref{Fig:4-2} shows as a result the composition of o-Fe$_{1-y}$Co$_y$Se$_2$ and c-Fe$_{1-z}$Co$_{z}$Se$_2$ as a function of $x_\textnormal{avg.}$. Notably, $y$ appears to saturate around 0.6 on increasing $x_\textnormal{avg.}$.

\section{DFT results}

\begin{figure}
    \centering
    \includegraphics[width=0.5\textwidth]{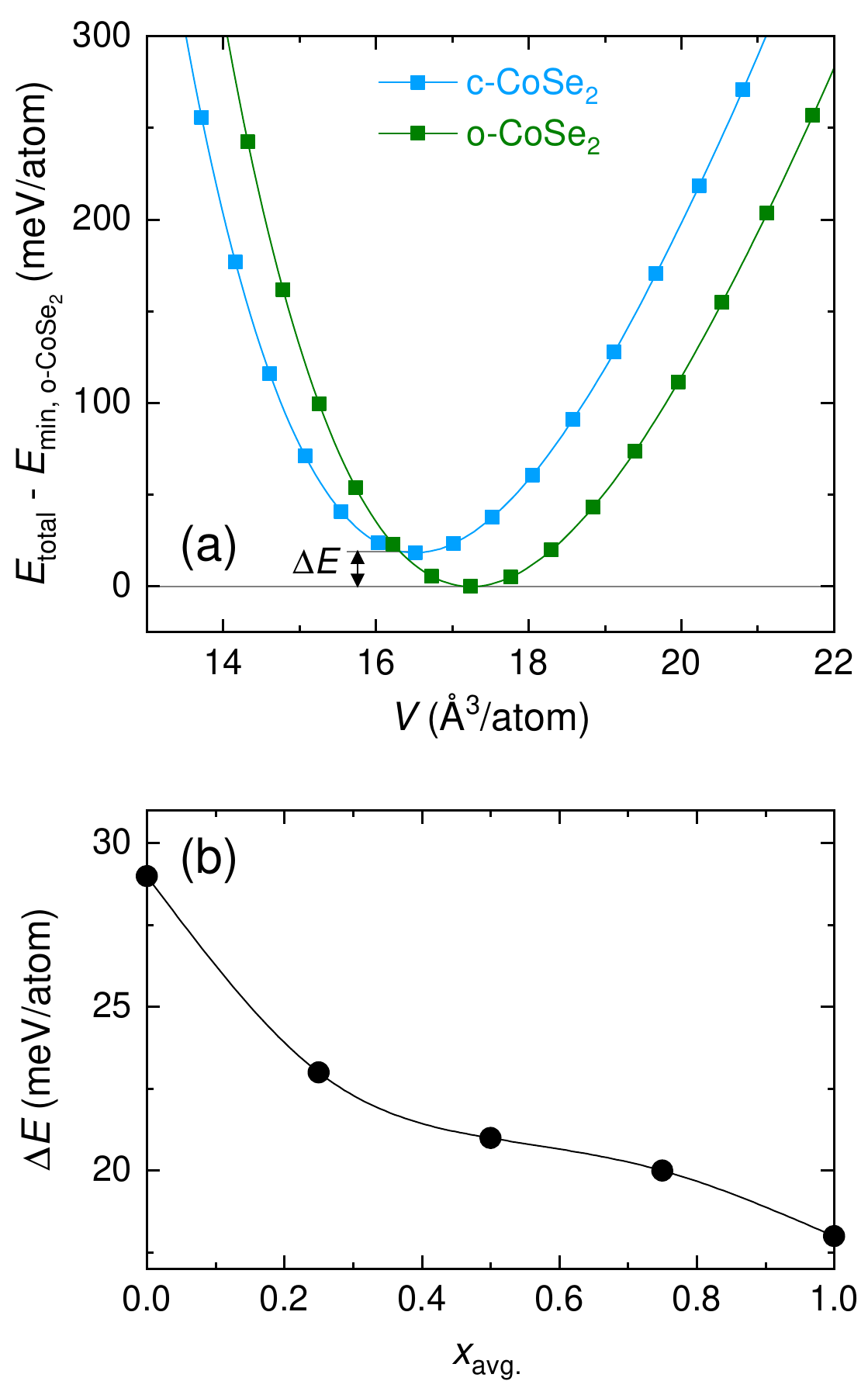}
    \caption{(a) Energy-volume curves of o-CoSe$_2$ and c-CoSe$_2$ obtained from DFT. (b) Energy difference $\Delta E$ between the minimum total energies of the orthorhombic and cubic phase as a function of the relative Co content $ x_\textnormal{avg.}$. Lines are a guide to the eye only.}
     \label{Fig:EV}
\end{figure}

The relative stability of cubic and orthorhombic o-Fe$_{1-y}$Co$_y$Se$_2$ and c-Fe$_{1-z}$Co$_{z}$Se$_2$ was investigated by DFT, see Fig.~\ref{Fig:EV}. Interestingly, the DFT calculations predict the orthorhombic phase as the ground state for all compositions. For pure CoSe$_2$, the energy difference between both phases is, however, only 18 meV/atom, corresponding to a thermal energy of 209 K. Furthermore, the energy difference is sensitive to the volume. Importantly, the energy barrier for the transition between orthorhombic and cubic phase, which proceeds through a reorientation of Se-bonds \cite{Zhang_2023}, is about 67 meV per atom.
With decreasing Co content, the energy difference between the two phases increases and reaches 29 meV/atom for FeSe$_2$.

The observed changes of lattice parameters with Co content can also be seen in DFT, see Fig.~\ref{Fig:3}. The deviation of lattice constants of all phases and Co contents predicted by theory and found in experiments is below 0.8\%. Overall, these findings are in agreement with previous studies \cite{Gavhane_2021, Gudelli}.

\begin{figure}
    \centering
    \includegraphics[width=0.5\textwidth]{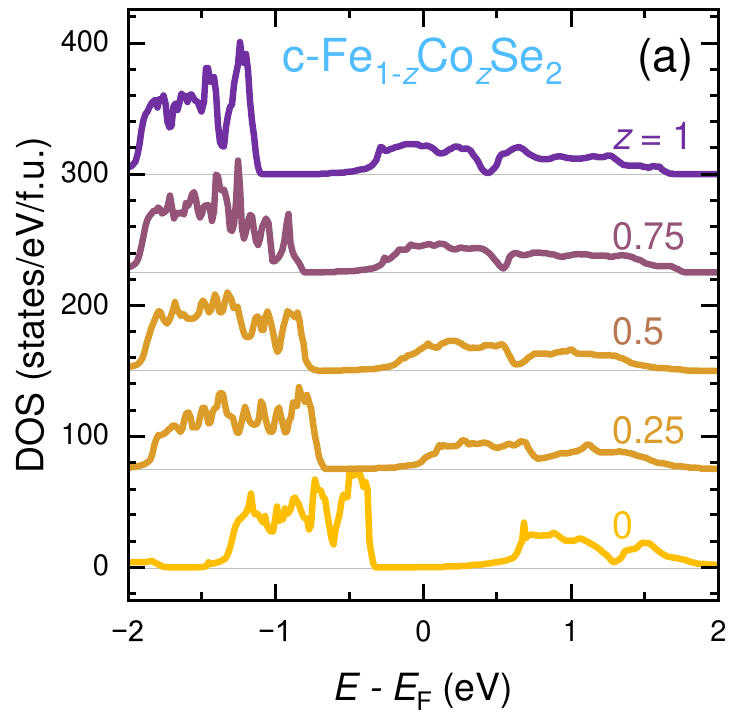}
     \includegraphics[width=0.5\textwidth]{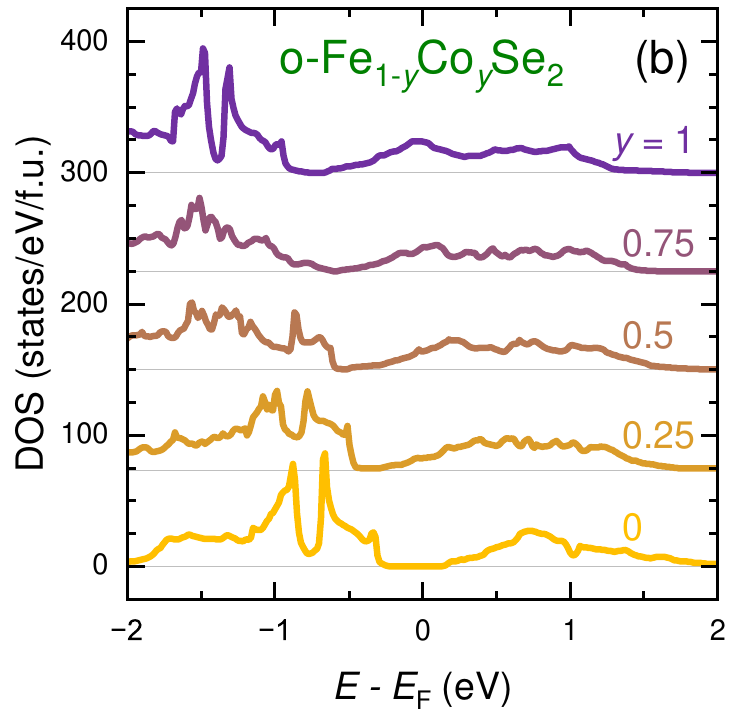}
    \caption{Total Density of states (DOS) per formula unit of  Fe$_{1-x}$Co$_{x}$Se${_2}$, offset by $75$ states per formula unit in the (a) cubic pyrite structure and (b) orthorhombic marcasite structure.
    \label{fig:dos}
    }
\end{figure}

Figure \ref{fig:dos} shows the evolution of the electronic densities of states (DOS) with increasing Co content for both polymorphs. With the selected technical details without Hubbard-$U$ corrections, an indirect band gap of only 0.3 eV, is found for orthorhombic FeSe$_2$, in comparison to 0.95 eV -- 1.1 eV found in experiment \cite{Kwon2007, Mars_2015} 
in agreement to early DFT predictions \cite{Yi2020}.
The valence band is predominantly composed of Co and Fe 3$d$-states. In the orthorhombic phase, these split to a  characteristic double-peak structure, with 3$d_{z^2}$-states close to the Fermi level due to the distortion of the metal-Se octahedra. 
The width of the valence band in orthorhombic CoSe$_2$ is larger compared to cubic CoSe$_2$. With decreasing Co content, the bandwidth systematically increases, indicating an enhanced degree of covalent bonding and corroborating the expected correlation with the energy lowering, see Fig.~\ref{Fig:EV}(b).

\section{Discussion}

Our detailed composition and synthesis-temperature dependent experimental results combined with DFT calculations give new insights into the formation of Fe$_{1-x}$Co$_x$Se$_2$.
We find that Fe$_{1-x}$Co$_x$Se$_2$ can be reproducibly synthesized via selenization even at temperatures as low as 250--430\textdegree\,C, much lower than previously reported selenization temperatures \cite{Hamdadou2002}. FeSe$_2$ and CoSe$_2$ indeed form at 250\textdegree\,C with complete reaction of the transition-metal films within 60 hours. 

The extreme composition CoSe$_2$ forms to 100\% or 94\% in the cubic phase when synthesized at 430\textdegree\,C or 350\textdegree\,C, respectively. The occurrence of the cubic phase is strongly reduced when the synthesis temperature is lowered to only 250\textdegree\,C. Here, only a fraction of 25\% of c-CoSe$_2$ is found and o-CoSe$_2$ dominates. On the Fe-rich side of the thin-film Fe$_{1-x}$Co$_x$Se$_2$ materials library, we observe the orthorhombic phase consistently for any synthesis temperature, in agreement with literature \cite{Kim2023}. Consistent with the experiment, our ab initio calculations neglecting temperature indeed show that the orthorhombic phase represents the ground state of CoSe$_2$. This tendency becomes more pronounced with decreasing Co content $x$ in Fe$_{1-x}$Co$_x$Se$_2$, as the energy difference between the orthorhombic and cubic phases increases. Overall, this shows that the orthorhombic structure is indeed the equilibrium phase of CoSe$_2$ at temperatures $T\lesssim 250$\textdegree\,C, and the equilibrium phase of Fe$_{1-x}$Co$_x$Se$_2$ in the entire composition range.  

We observe the coexistence of orthorhombic and cubic phases between  $x_\textnormal{avg.}=0.6-1$ for the two higher synthesis temperatures of 350\textdegree C and 430\textdegree C. For example, in the film synthesized at 430\textdegree\,C at composition $x_\textnormal{avg.}=0.8$, the orthorhombic phase o-Fe$_{1-y}$Co$_y$Se$_2$ with $y=0.65$ and the cubic phase c-Fe$_{1-\textit{z}}$Co$_{\textit{z}}$Se$_2$ with $z=0.87$ are found. The cubic phase is thus Co-enriched, which can be linked to its higher stability on the Co-rich side. Notably, both phases are well-formed as shown by the narrow Bragg peaks in XRD. This means that the transition-metal atoms are sufficiently mobile to move and form grains of the different phases with different Co contents.   

 It was suggested that a large size difference between the transition-metal atom and chalcogen atom in a pyrite-marcasite compound favors the formation of the orthorhombic marcasite structure, even when an electron counting rule suggests that the cubic phase should be more stable \cite{Witthaut2025}. Indeed, the proposed electron counting rule suggests the cubic structure for CoSe$_2$, but the large size difference between Co and Se seems to be sufficient to slightly favor the orthorhomic structure. 
 When synthesized at higher temperatures, the CoSe$_2$ polymorph forms in the cubic structure with higher symmetry. A phase transformation to the orthorhombic structure on lowering the temperature is most likely kinetically inhibited due to the large energy barrier for this transition.

\section{Summary and conclusions}

We have synthesized materials libraries of Fe$_{1-x}$Co$_x$Se$_2$ at different low temperatures to gain insight in the formation of the different polymorphs. We find that the orthorhombic marcasite polymorph can be stabilized to higher Co content $x$ when the synthesis temperatures are low. At a synthesis temperature of 250\textdegree C, the marcasite phase is the majority phase in the entire composition range. Density functional theory calculations show that the two polymorphs are extremely close in energy and the material is at the brink of a marcasite-pyrite transformation.

\section{Acknowledgements}
The authors acknowledge the use of the facilities of the center for interface-dominated high-performance materials (ZGH) at the Ruhr University Bochum. This work was partially supported by the Deutsche Forschungsgemeinschaft (DFG) under CRC/TRR 288 (Project A02).


\bibliography{Literature_FeCo2,Susanne}

\end{document}